\documentclass[12pt,preprint]{aastex}         

\usepackage{graphicx,epsfig,amssymb,layout,verbatim,rotating,calc,mathrsfs,epstopdf}
\usepackage[sumlimits,intlimits,namelimits]{amsmath}
\usepackage{graphics}
\usepackage{rotate}
\PassOptionsToPackage{linktocpage}{hyperref}
\usepackage{natbib}
\def\mnras{MNRAS}

\def\apj{ApJ}
\def\apjl{ApJL}

\def\apjs{ApJSS}
\def\be{\begin{equation}}
\def\ee{\end{equation}}

\def\tt{\mbox{\boldmath $\theta $}}

\def\lsim{\lower 2pt \hbox{$\, \buildrel {\scriptstyle <}\over
         {\scriptstyle \sim}\,$}}

\begin{document}
\newcommand{\figureout}[3]{\psfig{figure=#1,width=5.5in,angle=#2}
   \figcaption{#3} }

\title{GAMMA-RAY LIGHT CURVES FROM PULSAR MAGNETOSPHERES WITH FINITE CONDUCTIVITY}

\author{Constantinos Kalapotharakos\altaffilmark{1,2}, Alice K. Harding\altaffilmark{2}, Demosthenes Kazanas\altaffilmark{2} \\
and Ioannis Contopoulos\altaffilmark{3}}
\affil{$^1$University of Maryland, College Park (UMDCP/CRESST), College Park, MD 20742, USA;\\
$^2$Astrophysics Science Division, NASA/Goddard Space Flight Center, Greenbelt, MD 20771, USA;\\
$^3$Research Center for Astronomy and Applied Mathematics, Academy
of Athens, Athens 11527, Greece;\\
constantinos.kalapotharakos@nasa.gov}

\begin{abstract}

We investigate the shapes of $\gamma-$ray pulsar light curves using
3D pulsar magnetosphere models of finite conductivity. These models,
covering the entire spectrum of solutions between vacuum and
force-free magnetospheres, for the first time afford mapping the GeV
emission of more realistic, dissipative pulsar magnetospheres. To
this end we generate model light curves following two different
approaches: (a) We employ the emission patterns of the slot and
outer gap models in the field geometries of magnetospheres with
different conductivity $\sigma$. (b) We define realistic
trajectories of radiating particles in magnetospheres of different
$\sigma$ and compute their Lorentz factor under the influence of
magnetospheric electric fields and curvature radiation-reaction;
with these at hand we then calculate the emitted radiation
intensity. The light curves resulting from these prescriptions are
quite sensitive to the value of $\sigma$, especially in the second
approach. While still not self-consistent, these results are a step
forward in understanding the physics of pulsar $\gamma-$radiation.
\end{abstract}

\keywords{pulsars: general---stars: neutron---Gamma rays: stars}

\pagebreak

\section{INTRODUCTION}

The {\em{Fermi}} Gamma-Ray Space Telescope has had a major impact on
our understanding of pulsar physics with the discovery of over 100
$\gamma-$ray pulsars comprising three populations: young radio-loud
pulsars, young radio-quiet pulsars and millisecond pulsars
\citep{abdoetal2010}. Studying $\gamma-$ray pulsars with such a
broad range of underlying physical parameters offers an opportunity
of deeper understanding the physics underlying the pulsar
$\gamma-$ray emission and magnetic field geometry. A major issue
resolved early in the {\em Fermi} mission was the site of the pulsar
high-energy (GeV) emission. The cutoff of the Vela pulsar
phase-averaged spectrum measured by the Fermi Large Area Telescope
(LAT) \citep{abdoetal2009} ruled out at high significance the
super-exponential shape of magnetic pair production attenuation in
polar cap cascades and established the location of the $\gamma-$ray
emission and particle acceleration in the outer magnetosphere.

Present models for pulsar high-energy emission assume a vacuum
retarded dipole (VRD) \citep{Deutsch1955} field geometry, which is
expedient but fundamentally inconsistent. Such models nevertheless
have had some success in modeling Fermi pulsar light curves (LCs).
Outer Gap (OG) \citep{chr1986,romyad1995,hir2008} and Slot Gap (SG)
\citep{mushar2004} models both derive regions of $E_\parallel$
bordering the last open field-lines extending to the light-cylinder.
The particles accelerating in these `gaps' produce curvature
radiation (CR) and inverse-Compton emission, with their Lorentz
factors limited by CR reaction forces that balance $E_\parallel$.
The pattern of emission on the sky shows caustics that form on the
trailing edge of the open field region of each magnetic pole as
phase shifts due to dipole geometry, time-of-flight and aberration
nearly cancel, and photons from a large range of altitudes arrive in
phase \citep{morini1983,dykrud2003}. Observers viewing at angles
crossing the caustics will see one or two narrow peaks that resemble
the $\gamma-$ray LCs seen by Fermi. Because the GeV emission in
these models takes place in a region near the pulsar last open
field-lines, the shape of the model LCs is sensitive to, and thus a
good diagnostic of, the geometry of the pulsar magnetosphere near
the light-cylinder.

Fortunately, this can now be addressed in detail, thanks to recent
advances in numerical simulation of pulsar magnetospheres that model
the high-altitude field structure critical to the high-energy
emission. The global structure of realistic pulsar magnetospheres
remains an unsolved problem. Until recently, pulsar LC modeling has
employed the magnetospheric geometries of VRD and
force-free-electrodynamics (FFE)
\citep{ckf1999,spitkovsky2006,timokhin2006,kc2009}. The effects of
acceleration fields \citep{hir2007,hir2008} and open-zone currents
\citep{romwat2010} on the LC have been explored, but these models
are not fully self-consistent. In all these models, the sweepback of
the magnetic field-lines near the light-cylinder, due to retardation
and currents, causes an offset of the polar cap (PC) (and of the
entire magnetosphere) in the direction opposite to the rotation,
which can affect the $\gamma-$ray LCs. \cite{baispi2010} modeled
$\gamma-$ray LCs in FFE field geometry injecting photons along
tangents to the field direction of the separatrix. \cite{ck2010}
injected photons only in regions of the FFE magnetosphere with
$J/\rho{c}=1$, where $J$ is the current and $\rho$ is the local
charge density, as in these regions the electron velocities are
expected to be sufficiently close to $c$ to lead to GeV photon
production. More recently, \cite{haretal2011} assumed SG geometry to
produce model LCs for the VRD and FFE magnetospheres, concluding
that the VRD geometry provides better fits to the observed LCs than
the FFE geometry. The FFE models present larger field-line sweepback
and consequently the corresponding LCs have larger phase-lags
relative to the radio pulse which is not consistent with that of
observed $Fermi$ LCs.

Most recently, resistive magnetosphere models have appeared in the
literature \citep[Kalapotharakos et al. 2012 (K12,
hereafter);][]{lst2012}, that drop the ideal-MHD requirement in
favor of an Ohms' Law that relates the current to the $E$ and $B$
fields through a finite conductivity. These simulations reveal a
range of magnetic field structures, current distributions and
spin-down power that lie between the VRD and the FFE solution. Most
importantly, models of finite conductivity possess regions of
$E_\parallel$ which are potential locations of high-energy emission.

In this Letter, we explore for the first time the high-energy
emission that is generated in resistive magnetospheres, using
their magnetic field structure and $E_\parallel$ to produce
$\gamma-$ray LCs following two different approaches. First, we
assume the emission geometry of the SG and OG models and compute the
resulting LCs for different values of conductivity to compare with
those of the VRD and FFE geometries. In the second approach, we
define approximate particle trajectories, and we calculate the
corresponding energies (including radiation losses) and CR emission.
While neither of these approaches is self-consistent, in that
particle motions should affect the fields, they are an important
first step in relating the field structure and acceleration dictated
by global magnetosphere solutions to observations.

\section{LIGHT CURVE MODELING}

\subsection{Pulsar Magnetosphere Models with Finite Conductivity}

Dissipative magnetospheres are necessary for modeling pulsar LCs
considering that neither the VRD nor the FFE solutions are
compatible with the emission of radiation: The VRD solutions provide
maximum accelerating field, $E_{\parallel}$, but no charges
($\rho=0$), while the FFE solutions have a sufficiently large number
of particles to guarantee the nulling of $E_{\parallel}$. Herein, we
use the dissipative solutions presented recently by K12, which use a
phenomenological conductivity $\sigma$ ({\em in lieu} of
microphysical processes) to relate the current density $\mathbf{J}$
to the fields $\mathbf{E,~B}$. Although these solutions are still
not self-consistent, they are an improvement over those of VRD and
FFE because, besides the global field geometry, they also provide
the distribution of $E_\parallel$, a quantity necessary to compute
the acceleration of radiating charges.

In the next sections we present model LCs using the structure of
magnetic and electric fields provided by dissipative solutions of
the perpendicular rotator ($\alpha=90^{\circ}$) which span the
entire solution space from VRD to FFE. These solutions are presented
in detail in K12 and have been produced adopting a very simple
prescription for the current density
\begin{equation}\label{current}
    \mathbf{J}=c\rho\frac{\mathbf{E}\times\mathbf{B}}{B^2}+\sigma
    \mathbf{E_{\parallel}}\;
\end{equation}
In this case the current density consists of two components, namely
a drift current and a component parallel to the magnetic field.
While some simulations explored models with a spatially dependent
$\sigma$, in this paper we will use only those with constant
$\sigma$. As $\sigma$ goes from 0 to $\infty$ the corresponding
solution ranges from VRD to FFE. The solutions we consider here
correspond to $\sigma\approx 0.08, 1.5$ and $24\Omega$, where
$\Omega$ is the angular frequency of the star. The field structure
of the solutions for $\sigma\approx1.5\Omega$ and
$\sigma\approx24\Omega$ are shown in the last rows of figures 4 and
3 of K12, respectively.

\subsection{Geometric Approach}

A simple method of generating pulsar LCs, used in many previous
studies \citep[e.g][]{dykhar2004,wrwj2009,vhg2009}, adopts the
geometry of physical emission models that have computed the shape of
accelerator gaps with assumed field structure and sources of charge.
This method can directly compare LCs from magnetospheres having
finite conductivity with LCs in VRD and FFE magnetospheres. To
explore how the magnetic field structure and offset PCs influence
$\gamma-$ray pulsar LCs, we have generated model LCs using a
geometrical version of the SG and OG models
\citep[e.g.][]{dykhar2004,wrwj2009,vhg2009}. The SG has its origin
in PC pair cascades that screen the accelerating parallel electric
field $E_{\parallel}$ over most of the open field except in narrow
gaps along the last open field-lines \citep{arons1983}. The
electrons accelerate and radiate from the neutron star (NS) surface
to high-altitude, and emission occurs throughout the volume of the
gap \citep{mushar2004,haretal2008}. The OG is a vacuum gap that also
forms adjacent to the last open field-line, above the null charge
surface where the corotation charge \citep{goljul1969} changes sign.
The gap width is determined by the screening of $E_\parallel$ by
pair cascades and emission occurs in a thin region along the gap
inner edge \citep{wanetal2010}.

Components of the magnetic field are determined from analytic
expressions for the VRD \citep[see][]{dykhar2004} and interpolated
from numerical simulations for FFE and resistive magnetospheres. The
open field boundary on the NS surface (the PC rim) was determined
via bisection in magnetic colatitude at fixed azimuth values. Open
Volume radial and azimuthal Coordinates (OVC) were then defined
inside the open volume of each solution \citep{dykhar2004}. We
assume that particles travel from the NS surface along open
field-lines in OVCs and emit radiation tangent to field-lines,
uniformly, in the corotating frame (CF).  We assume that emission is
also uniform across a SG of width $w=0.05$, as a fraction of the
open volume, on field-lines originating between $r_{\rm min}=0.95$
and $r_{\rm max}=1.0$ on the PC (in units of PC radius) and in a
thin layer at $r_{\rm{min}}\simeq{r_{\rm{max}}=0.95}$ in the OG. The
minimum and maximum spherical radii of emission are assumed to be
the NS surface and $R_{\rm{max}}=1.2R_{\rm{LC}}$, limited by a
maximum cylindrical radius of $R_{\rm max}^{\rm cyl}=0.95R_{\rm
LC}$, for the SG, and the null surface and $R_{\rm max}=1.5 R_{\rm
LC}$, limited by $R_{\rm max}^{\rm cyl}=0.97R_{\rm LC}$, for the OG.
The photon direction is assumed to be tangent to the magnetic field
in the CF, obtained through a Lorentz transformation from the
inertial observer's frame (IOF) \citep{baispi2010}. The emission
direction is then transformed to the IOF (aberration), time-delays
are added and the emission is accumulated in sky-maps in viewing
angle $\zeta$ and phase $\phi$ with respect to the pulsar rotation
axis. LCs are then obtained as slices through these maps at constant
$\zeta$.

\subsection{Particle Trajectory Approach}

An altogether different approach to produce LCs that takes
approximately into account the electric fields present in the
specific magnetospheric solution is the following: Since we
anticipate the velocity of any particle of the magnetosphere to be
very close to $c$, we decompose its motion into a drift component
and one parallel to the magnetic field. Thus, one can write for the
particle velocity
\begin{equation}\label{veloc}
    \mathbf{u}=\frac{\mathbf{E}\times\mathbf{B}}{B^2}c+f c\frac{\mathbf{B}}{B}
\end{equation}
The sign and the absolute value of the factor $f$ is chosen so that
the motion of the particle be outward and the total modulus of the
velocity $u$ be $c$. This trajectory determination is similar to
those of \cite{ck2010} and \cite{baispi2010}. Under this assumption
we calculate the trajectories passing through each magnetospheric
point inside a central cube of edge $3R_{\rm{LC}}$ considering that
in the dissipative solutions the particles do not follow the
field-lines in the CF. So,instead of open and closed field-lines we
determine open and closed trajectories depending on whether they
reach (or not) $2R_{\rm{LC}}$, assuming that the radiating particles
follow only open trajectories\footnote{Our field structures have
$E_{\parallel}$ in the closed zone too; we believe that this is due
to approximations in our field computations and therefore we
restrict radiation by particles accelerated only by the
$E_{\parallel}$ of the open zone.}. This particle trajectory
determination allows also the calculation of the local radius of
curvature $R_{cr}$ at each point of the magnetosphere. Moreover,
assuming that each particle starts at the stellar surface with a
small $\gamma$-value $(\gamma\lesssim100)$ we can calculate its
Lorentz factor $\gamma$ along its trajectory from
\begin{equation}\label{dgamma}
    \frac{d\gamma}{dt}= f\frac{q_e c E_\parallel}{m_e
    c^2}-\frac{2}{3}\frac{q_e^2 \gamma^4}{R_{cr}^2 m_e c}
\end{equation}
where $q_e$ and $m_e$ are the electron charge and rest-mass
respectively. This approach allows us to have all the information
needed to calculate the CR intensity contributed by each point of
the magnetosphere and so the corresponding sky-maps and the LCs (see
Section 2.2).

\section{RESULTS}

Figure 1 shows geometric LCs for SG (left-hand column) and OG
(right-hand column) emission in VRD (black) and FFE (purple)
solutions, and in resistive magnetospheres with $\sigma=0.08\Omega$
(red), and $1.5\Omega$ (green) for a range of observer angles
$\zeta$. The resistive solutions of lowest $\sigma$ are closest to
the VRD and indeed, the computed LCs look very similar. However,
there is a shift to larger phase and a slight broadening of the
peaks of the $\sigma=0.08\Omega$ LCs. As $\sigma$ increases, in both
the SG and OG cases, the peaks are shifted even more to larger phase
and the broadening is more pronounced, with the highest
$\sigma$-value being closest to the FFE solutions. Indeed, their LCs
look very similar, with only a slight shift in phase of the peaks
but no further broadening. Overall, there is a distinct progression
in the LC shapes as conductivity increases. The VRD LCs have the
narrowest peaks and the smallest phase-lag from the magnetic pole
(phase$=0$), with peak-width and phase-lag systematically increasing
with conductivity.

The LC changes with conductivity result from changes in magnetic
field structure.  Magnetospheres with low $\sigma$ are ``stiffer"
and thus have less sweptback field-lines and smaller open field
volume, while those with higher $\sigma$ have more sweepback
\citep{baispi2010}. The increase of sweepback with $\sigma$ produces
a larger shift of the PC which causes the larger phase-lag of the LC
peaks. The increase in open volume of magnetospheres with large
$\sigma$ also causes the increase in peak-width, since the gap
widths are assumed to be a fraction of the open volume. Since
peak-width and phase-lag are measurable characteristics of observed
$\gamma-$ray pulsar LCs, this study shows that they could
potentially be an important diagnostic of magnetospheric
conductivity. Comparison of geometric LCs in VRD and FFE
magnetospheres has already indicated that VRD provides a better
match to observed LCs \citep{haretal2011}.

In Figure~\ref{figlc2} we plot the LCs for solutions corresponding
to three different values of $\sigma$ taking into account the
physical properties provided by each solution as we describe in
Section 2.3. The red, green and blue color correspond to
$\sigma=0.08\Omega,~1.5\Omega,~24\Omega$, respectively. For these
LCs we assume that emission occurs only along all open trajectories
(i.e. those that reach at least up to $2R_{\rm{LC}}$). The emission
is considered to be due to CR and is always proportional to
$\gamma^4 R_{\rm cr}^{-2}$. The $\gamma$-value is derived by
Eq.~\ref{dgamma}. The total emissivity can also be weighted by the
local charge density $\rho$ (left-hand column of
Figure~\ref{figlc2}). The general feature is that the broadest
(narrowest) pulses seem to be those corresponding to the middle
(high) $\sigma$-value $\sigma=1.5\Omega$ ($\sigma=24\Omega$).
However, the LCs for $\sigma=24\Omega$ and for the low
$\zeta$-values exhibit high off-pulse emission. This effect
decreases when the charge density $\rho$ weighting is included.
Moreover, near $\zeta=90^{\circ}$, the middle $\sigma$-value
($\sigma=1.5\Omega$) pulses are weak\footnote{This is not shown in
the normalized LCs of Figure~\ref{figlc2} but only in the
corresponding sky-maps.}, double and narrow. The general trend for
the $\sigma=0.08\Omega$ and $\sigma=1.5\Omega$ solutions is that the
phase-lag of the pulses with respect to the magnetic poles
($\rm{phase}=0$) decreases with $\zeta$ although there are counter
examples. These phase-lags start from values higher than 0.25 (for
$\zeta=45^{\circ}$) and only for $\zeta$ near $90^{\circ}$ can reach
close to 0.1-0.15. For the high $\sigma$-value ($\sigma=24\Omega$)
the corresponding phase-lags seem to be near the value 0.25 for most
of the cases implying a non-monotonic behavior with $\sigma$.

We checked also the assumption that the emitting particles are
mostly those that follow the high-voltage trajectories\footnote{The
voltage is given by $\int\mathbf{E\cdot{dl}}$ along a trajectory}.
This approach is similar to the geometric one (Section 2.2) in the
sense that it is supposed that only a part of the magnetosphere
contributes to the emission. The difference here is that the active
region is traced by the high-voltage trajectories.
Figure~\ref{figlc2b} is similar to Figure~\ref{figlc2} but only the
10\% of the highest-voltage trajectories are considered to emit. In
this case we observe narrower pulses. However, only for the middle
$\sigma$-value ($\sigma=1.5\Omega$) we have pulses corresponding to
small phase-lags. We note also that the low $\sigma$-value
($\sigma=0.08\Omega$) is observable only from a relatively narrow
range of $\zeta$ ($\zeta\approx45^{\circ}-60^{\circ}$).

We derived also the LCs considering each point's emissivity
$\propto\rho_{s}\gamma^4R_{cr}^{-2}$ where $\rho_{s}$ is the charge
density on the star surface at the point where each particle starts
its journey. In this case the results are similar to those of the
right-hand columns of Figures~\ref{figlc2},~\ref{figlc2b}.

\begin{figure}
\centering
\includegraphics[width=14cm]{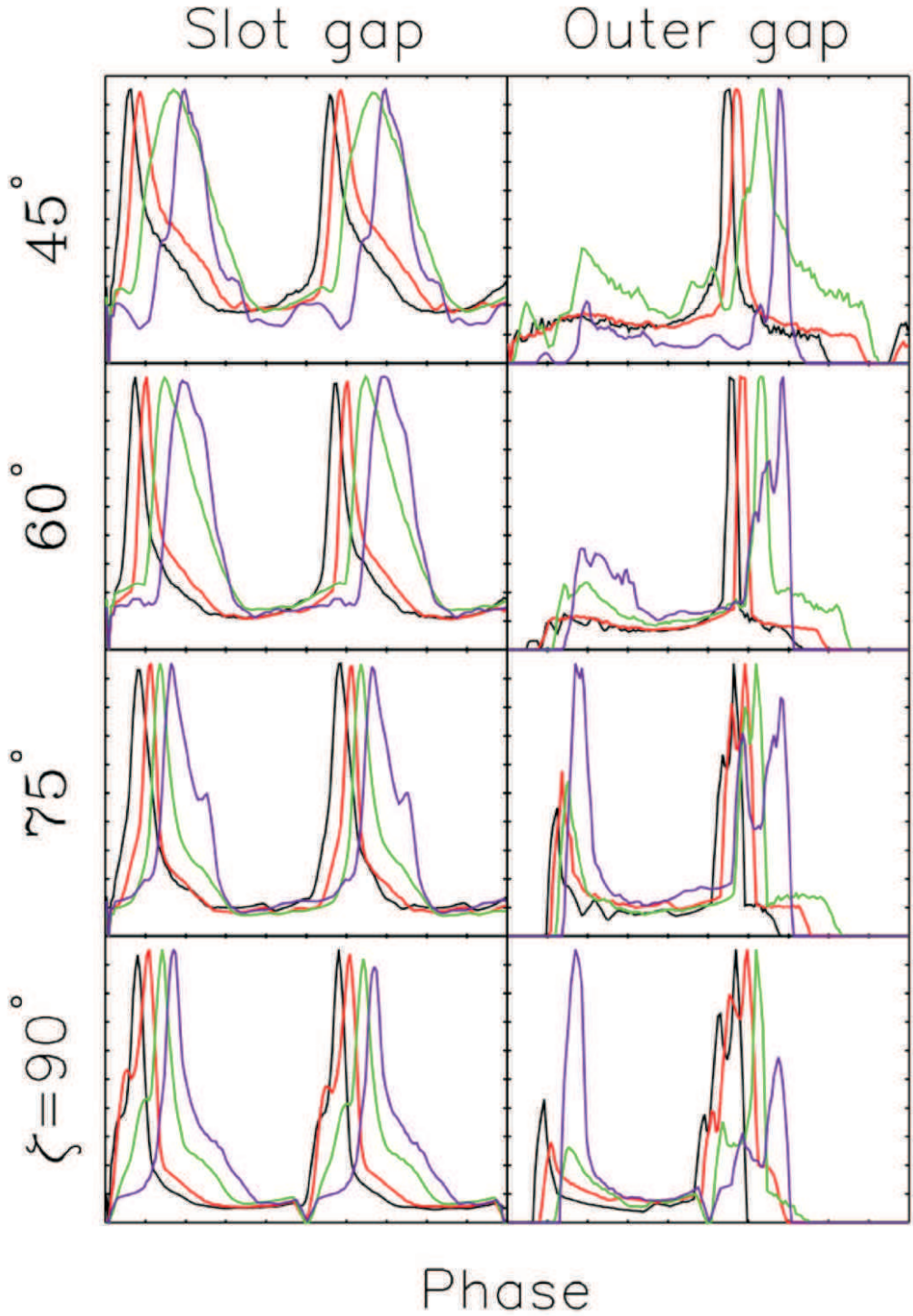}
\caption{Geometric LCs for slot gap (left) and outer gap (right)
emission in VRD (black), FFE (purple) solutions, and resistive
magnetospheres with $\sigma=0.08\Omega$ (red) and $\sigma=1.5\Omega$
(green) for pulsar inclination angle $\alpha=90^\circ$.}    
\label{figlc1}
\end{figure}

\begin{figure}
 \centering
  \includegraphics[width=10cm]{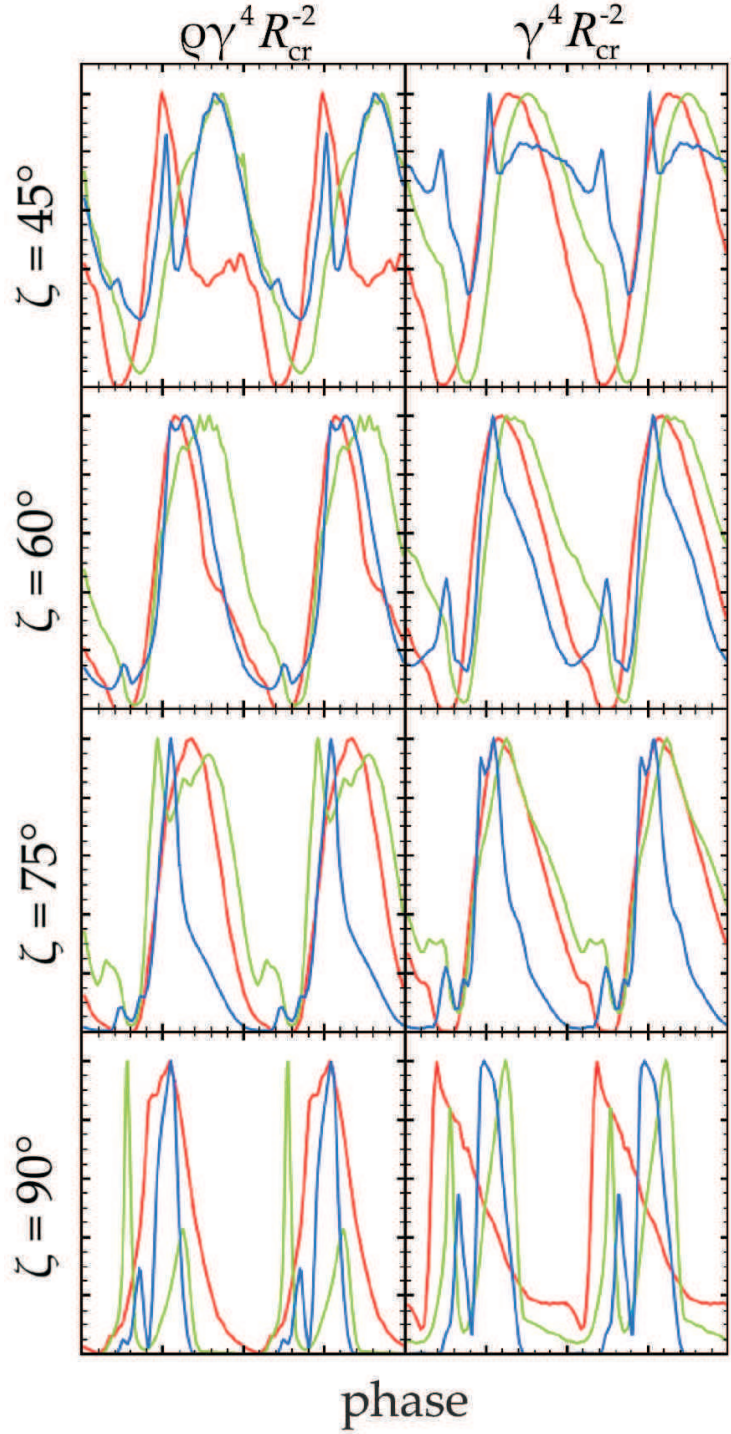}\\
\caption{The LCs corresponding to the trajectory approach. The
assumed emissivity at each point of the magnetosphere is indicated
in the figure. The red, green and blue colored lines correspond to
$\sigma=0.08\Omega,~1.5\Omega~\text{and}~24\Omega$, respectively.}
\label{figlc2}
\end{figure}

\begin{figure}
 \centering
  \includegraphics[width=10.7cm]{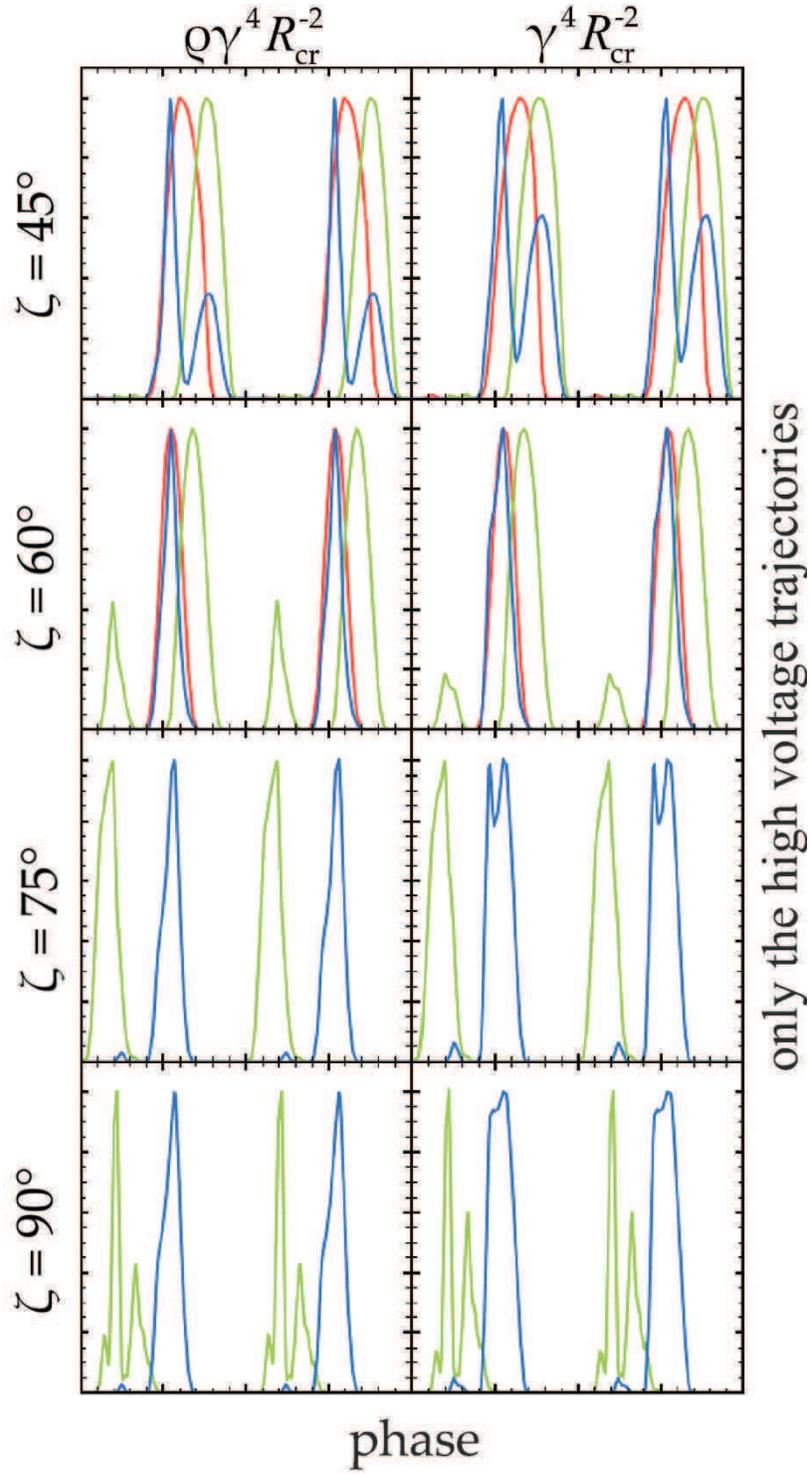}\\
\caption{Similar to Figure~\ref{figlc2} but only the 10\%
highest-voltage trajectories are assumed to emit.} \label{figlc2b}
\end{figure}

\begin{figure}
\centering
  \includegraphics[width=\textwidth]{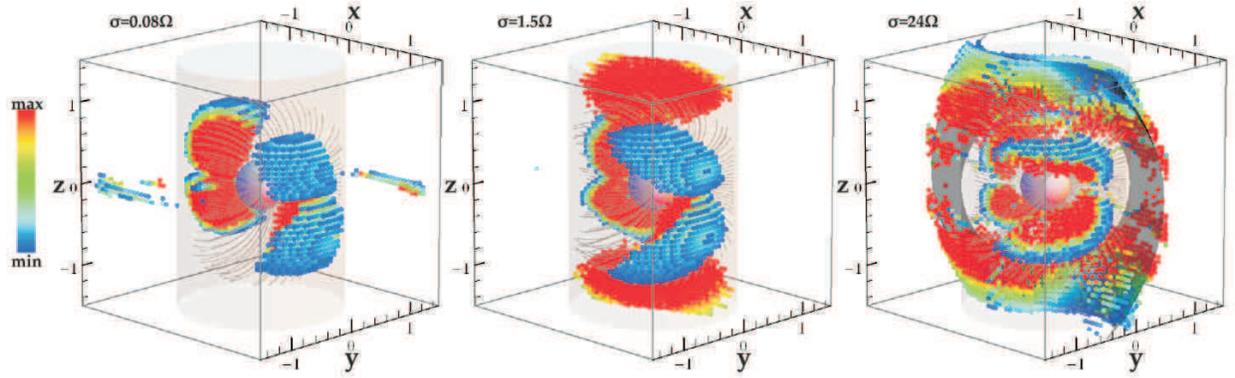}\\
\caption{The regions of the magnetospheres that produce the peaks of
the pulses shown in the right-hand column of Figure~\ref{figlc2}.
The color scale indicates the corresponding emissivity. For
$\sigma=24\Omega$ a significant part of the emission comes from a
region near the current-sheet outside the light-cylinder (gray
surface in the right-hand panel).} \label{fig3D}
\end{figure}

In Figure~\ref{fig3D} we plot the points in the 3D magnetosphere
that produce the pulses shown in the right-hand column of
Figure~\ref{figlc2}. In each of these cases we have identified the
phases of the observed pulses for all $\zeta>30^{\circ}$ and we
located the points that contribute over 90\% of the corresponding
emission. The emissivity of each point is represented by the
indicated color scale. In these figures we have also plotted the
portion of the magnetic field-lines (in gray) that contribute to the
emission in the SG model presented in Figure~\ref{figlc1}. We see
that there is always a blob of points over the polar caps
contributing to the observed pulses. However, as we go toward high
$\sigma$-values the volume of these blobs decreases and new points
from the outer magnetosphere are added. For $\sigma=0.08\Omega$ and
$\sigma=1.5\Omega$ the inner parts of the blobs coincide with only a
subset of the SG lines. This subset increases with $\sigma$ and at
$\sigma=24\Omega$ the largest part of the SG lines seem to follow
the colored points. Beyond the light-cylinder, this region coincides
with the current-sheet (Figure~4). This supports the result of
\citep{ck2010} where a significant part of the emission is produced
near the current-sheet. The changes of the form of the effective
emitting regions, described above, make the sky-maps evolution with
$\sigma$ complex. This makes the LCs
(Figures~\ref{figlc2},~\ref{figlc2b}) evolve in a non-monotonic
manner with increasing $\sigma$. However, inspection of the full
sky-maps shows smooth evolution with increasing conductivity.

Figures~\ref{figlc1}-\ref{fig3D} show that the geometric LCs are
less sensitive to $\sigma$ at high $\sigma$-values than those of the
trajectory approach. This indicates that the $E_{\parallel}$
distribution changes more significantly than the magnetic field
structure at high $\sigma$.

\section{DISCUSSION}

We have presented a first implementation of dissipative
magnetospheres to model pulsar emission for direct comparison with
the fast accumulating pulsar phenomenology. We have concentrated
here on the pulsar $\gamma-$ray LCs, however our models could also
address spectral features. With LC modeling alone, we have just
scratched the surface of the problem, since we have used only one of
the prescriptions discussed in K12 and only one inclination angle
$\alpha=90^{\circ}$. A more complex behavior is expected for a
variety of prescriptions and $\alpha$'s.

The main goal of this paper is to show that more realistic pulsar
magnetosphere models provide flexibility that allows meaningful
constraints on their parameters in direct comparison with
observations. While neither of the approaches employed are
self-consistent in that they take into account the effects of the
radiating particles on the magnetosphere itself, we find a
progression in the LC shape, with peak phase and width increasing
with $\sigma$ in the SG and OG models. Observed Fermi LAT LCs show
an inverse correlation between the peak separation $\Delta$ (in LCs
with two peaks) and phase-lag $\delta$ of the first peak relative to
the radio-peak (thought to be near $\rm{phase}=0$)
\citep{abdoetal2010}. The first-peak phase-lags ($\sim0.17-0.2$) of
high $\sigma$ model LCs are too large to account for the observed
$\delta$ of many LAT pulsars with $\Delta=0.5$. The LCs computed in
the particle trajectory approach can also produce narrow pulses,
despite the fact that they assume particle emission at every point
of the magnetosphere; however, their consistency with observations
may require specific values of $\zeta$, $\alpha$ and $\sigma$.
Independently of whether this is true or not, the final calculation
of the LCs depends also on the modulation by the local number of the
emitting particles, which is something that can be derived only in
fully self-consistent solutions. We plan to move in this direction
by introducing pair cascade mechanisms and the calculation of the
exact particle orbits (using the full equations of motion), taking
into account both CR and synchrotron losses, as well as
inverse-Compton radiation.

We expect that with the dissipative models at hand, comparison of
their model LC and spectral products with observations will allow us
to find specific magnetospheric prescriptions and model
parameters that provide the working physics of pulsar
magnetospheres. All these will be the subject of future work.

\acknowledgments

AKH acknowledges support from the NASA Astrophysics Theory and
Fundamental Physics Program and the $Fermi$ Guest Investigator
Program. We thank also the referee for his/her constructive
comments.

\newpage

\newpage

\newpage

\newpage

\clearpage
\newpage

\end{document}